\documentstyle[proceedings,epsfig]{crckapb} 

\def\vec#1{{\bf #1}}

\begin{opening}

   \title{The Hanle effect in 1D, 2D and 3D}

   \author{Rafael Manso Sainz}

   \author{Javier Trujillo Bueno}

   \institute{Instituto de Astrof\'\i sica de Canarias, E-38200\\
     La Laguna, (Tenerife), Spain}

\end{opening}

\runningtitle{The Hanle Effect}

\begin{document}

\begin{abstract}\footnotesize\ 
This paper\footnote{Published in 1999 in the book {\it Solar Polarization}, edited by K.N. Nagendra \& J.O. Stenflo. Kluwer Academic Publishers, 1999. (Astrophysics and Space Science Library ; Vol. 243), p. 143-156} 
addresses the problem of scattering line polarization
and the Hanle effect in one-dimensional (1D),
two-dimensional (2D) and three-dimensional (3D) media for the case
of a two-level model atom without lower-level polarization
and assuming complete frequency redistribution. The theoretical framework
chosen for its formulation is the QED theory of Landi Degl'Innocenti (1983),
which specifies the excitation state of the atoms in terms of the irreducible
tensor components of the atomic density matrix. The self-consistent
values of these density-matrix elements
is to be determined by solving jointly
the kinetic and radiative transfer equations for the Stokes parameters.
We show how to achieve this by generalizing
to Non-LTE polarization transfer
the Jacobi-based ALI method of Olson {\it et al.}
(1986) and the iterative schemes based on Gauss-Seidel iteration of
Trujillo Bueno and Fabiani Bendicho (1995). These methods 
essentially maintain the simplicity
of the $\Lambda-$iteration method, but their convergence rate is extremely
high. Finally, some 1D and 2D model calculations are presented
that illustrate the effect of horizontal 
atmospheric inhomogeneities on magnetic and non-magnetic resonance line polarization signals.


\end{abstract}

\section{Introduction}

The scattering line polarization, and its modification due to a
weak magnetic field ---such that the Zeeman splitting is negligible
compared with the line width
(the so called Hanle effect; Hanle, 1924)---,
sensitively depends on the {\it anisotropy} of the radiation field
and on the magnetic field vector geometry (Landi
Degl'Innocenti, 1985; Stenflo, 1994). The solar atmospheric plasma is spatially
inhomogeneous, with vertical and horizontal variations, 
not only in temperature, macroscopic velocity and density, but also in the orientation and intensity of the magnetic field
(see S\'anchez Almeida, 1999 for new insights in this respect). 
It is thus clear that, in order to fully exploit the 
Hanle effect as a diagnostic
tool for weak magnetic fields, we also need to address
the problem of resonance line polarization and the Hanle effect in 2D and 
3D media where the radiation field's anisotropy is
different from that corresponding to the currently-assumed
1D atmospheric models.

To this end, this contribution begins presenting a formulation of 
resonance line polarization and the Hanle effect
that we consider as the most suitable one for practical RT applications. 
It is based on the density-matrix theory for the generation and transfer
of polarized radiation (see Landi Degl'Innocenti, 1983; 1984; 1985). 
In this paper we consider the standard case of
a two-level model atom neglecting atomic polarization
in its lower level (i.e. it is assumed that the lower-level Zeeman sublevels
are equally populated and that there are no coherences among them).
The quantities whose {\it self-consistent} values are to be determined 
are the irreducible tensor components of the density matrix ($\rho^K_Q$),
which depend only on the spatial coordinates.
The statistical 
equilibrium (SE) and RT equations to be solved are
valid independently of whether we assume 1D, 2D or 3D geometries.

A summary of previous work done in the subject
of the numerical solution of Non-LTE polarization transfer
problems can be found in Trujillo Bueno and Manso Sainz (1999).
In this respect, we should mention the recent work of
Nagendra {\it et al.} (1998;
see also their contribution in these proceedings) 
where the Hanle effect in 1D is considered
using a different theoretical approach. 
For information concerning the numerical solution of
more general polarization transfer problems formulated
with the density-matrix theory see Trujillo Bueno (1999).

The outline of this paper is as follows. Section 2 presents the basic
equations for the case of a two level atom
without lower-level atomic polarization
and assuming complete frequency redistribution.
Section 3 shows how the very efficient iterative methods of solution 
investigated by Trujillo Bueno and Fabiani
Bendicho (1995) (Jacobi, Gauss-Seidel and successive over-relaxation) can
be suitably generalized to the problem of Non-LTE polarized radiative 
transfer in the Hanle effect regime. Finally, in Sect. 4 we present
the results of some illustrative 2D Hanle-effect calculations
for triplet lines 
and discuss the ensuing horizontal transfer effects.

\section{Basic Equations}

In scattering line polarization and Hanle effect problems, the quantum
interferences (or coherences)
between the magnetic sublevels of each atomic level
must be considered. In order to properly
take into account these effects, we work within the framework of the
polarization transfer theory based on the irreducible tensor components 
of the atomic density matrix (see Bommier and Sahal Br\'echot, 1978; 
Landi Degl'Innocenti, 1983). Since we are restricting
ourselves in this paper to the standard case
of a two-level model atom neglecting lower-level atomic polarization
only the $K=Q=0$ {\it irreducible tensor component} of the density matrix
suffices to completely describe the lower-level excitation. For instance,
if the lower level of total population $N_l$ has angular momentum $J_l=0$
one has that $\rho^0_0(l)=N_l/\sqrt{3}$.

However, with respect to the upper level we need 
to specify its total population ($\rho^0_0(u)$), its alignment
($\rho^2_0(u)$), and two complex quantities 
($\rho^2_1(u)$ and $\rho^2_2(u)$)
which take into account the coherences
between the sublevels of the upper level. For example,
for an upper level with $J_u=1$
\begin{eqnarray}
  \rho^0_0(u)=\frac{1}{\sqrt{3}} (N_1+N_0+N_{-1}) = \frac{1}{\sqrt{3}}
  N_{u}, 
\end{eqnarray}
\vspace{-0.3in}
\begin{eqnarray}
  \rho^2_0(u)=\frac{1}{\sqrt{6}} (N_1-2N_0+N_{-1}),
\end{eqnarray}
\vspace{-0.3in}
\begin{eqnarray}
  \rho^2_1(u)=-\frac{1}{\sqrt{2}} [\rho_{J_u}(1, 0)-\rho_{J_u}(0, -1)],
\end{eqnarray}
\vspace{-0.3in}
\begin{eqnarray}
  \rho^2_2(u)=\rho_{J_u}(1, -1),
\end{eqnarray}
where $N_i$ ($i=1,0,-1$) are the populations of the magnetic sublevels $M=0,\pm1$ of the upper level, and 
$\rho_{J}(M, M')\,=\,<\alpha\,J\,M|\rho|\alpha\,J\,M'>$ (with
$|\alpha\,J\,M>$ the eigenvectors of the atomic Hamiltonian)
are the elements of the density matrix in the standard representation
(see, e.g., Messiah, 1969). We point out that the {\it orientation}
components ($\rho^1_Q$) of the density matrix are zero because
we are assuming a static medium for which the radiation field
that illuminates its boundaries has no circular polarization
(see Landi Degl'Innocenti {\it et al.}, 1990).

For each irreducible {\it upper} level
tensor component $\rho^K_Q$ with $Q>0$, there
exists another spherical component with $Q<0$ related to it through
the conjugation property:  
\begin{equation}
  \rho^K_{-Q} = (-1)^Q [\rho^K_Q]^*,
\end{equation}
where the symbol ``$*$'' means complex conjugation. We can thus choose the
following linear
combinations as independent variables:
\begin{eqnarray}
  \tilde{\rho}^K_Q = \frac{1}{2} [\rho^K_Q +(-1)^Q \rho^K_{-Q}]\,=\,{\rm Re}\,[\rho^K_Q]\, , \,\,\,\,\,\,\,\,\, Q>0 ,\\
  \hat{\rho}^K_Q = \frac{1}{2i} [\rho^K_Q -(-1)^Q \rho^K_{-Q}]\,=\,{\rm Im}\,[\rho^K_Q]\, , \,\,\,\,\,\,\,\,\, Q>0 ,
\end{eqnarray}
where ``$i$'' is the imaginary unit. 
Finally, normalizing all these {\it upper} level {\it unknowns} to
the total atomic population of the lower level, we find the six {\it real}
unknowns of this 
problem: $\rho^0_0$, $\rho^2_0$, $\tilde{\rho}^2_1$, $\hat{\rho}^2_1$,
$\tilde{\rho}^2_2$ and $\hat{\rho}^2_2$ of the upper level.

In the QED polarization transfer theory of
Landi Degl'Innocenti (1983, 1984), the radiation field is
described by its spherical tensor components:
\begin{eqnarray}
  J^0_0=\int {\rm d}x \,\phi_x \oint \frac{{\rm d}
  \vec{\Omega}}{4\pi}\,I_{x \vec{\Omega}}
\end{eqnarray}
\vspace{-0.3in}
\begin{eqnarray}
  J^2_0=\int {\rm d}x \,\phi_x \oint \frac{{\rm d} \vec{\Omega}}{4\pi}
  \frac{1}{2\sqrt{2}} [(3\mu^2-1)I_{x \vec{\Omega}}+3(\mu^2-1)Q_{x
  \vec{\Omega}}]
\end{eqnarray}
\vspace{-0.3in}
\begin{eqnarray}
  J^2_1=\int {\rm d}x \,\phi_x \oint \frac{{\rm d} \vec{\Omega}}{4\pi}
  \frac{\sqrt{3}}{2} e^{i\chi} \sqrt{1-\mu^2}[-\mu(I_{x \vec{\Omega}}+Q_{x
  \vec{\Omega}})-iU_{x \vec{\Omega}}]
\end{eqnarray}
\vspace{-0.3in}
\begin{eqnarray}
  J^2_2=\int {\rm d}x \,\phi_x \oint \frac{{\rm d} \vec{\Omega}}{4\pi}
  \frac{\sqrt{3}}{2} e^{2i\chi}
  [\frac{1}{2}(1-\mu^2)I_{x \vec{\Omega}}-\frac{1}{2}(1+\mu^2)Q_{x
  \vec{\Omega}}-i\mu U_{x \vec{\Omega}}] 
\end{eqnarray}
where $I_{x\vec{\Omega}}$, $Q_{x\vec{\Omega}}$ and $U_{x\vec{\Omega}}$ are
the Stokes parameters relative to the direction $\vec{\Omega}$ specified by
the angles $\theta$ and $\chi$ defined as in Fig. (2b), $\mu=\cos \theta$, and
$\phi_x$ is the line profile, with $x$ the frequency measured from the line
center in units of the Doppler width.
The physical meaning of these expressions is quite simple. 
Note that $J^0_0$ is
the well-known frequency integrated mean intensity. The other three 
irreducible tensor components of the radiation field are
frequency and angular integrals of the three Stokes parameters, weighted by
some angle-dependent quantities, and 
by the line profile $\phi_x$. The radiation field tensor $J^2_0$
measures the degree of vertical-horizontal anisotropy: it is positive when
the radiation is predominantly vertical, negative if horizontal, and it
vanishes at the bottom of the atmosphere where the radiation field is
unpolarized and isotropic. Finally, the two other spherical tensor components
are complex, and they measure the breaking of the axial symmetry of the radiation
field through the azimuthal exponentials appearing inside the angular
integrals. Therefore, they are zero in axially-symmetric media like
1D plane-parallel atmospheres with a vertical magnetic field, 
or without any field at all.
The $J^K_Q$ components with $Q<0$ can be obtained through a
conjugation relation similar to 
the one stated in Eq. (5) for the atomic statistical tensors. With
equivalent definitions to those of Eqs. (6)-(7), we obtain the six real
quantities $J^0_0$,
$J^2_0$, $\tilde{J}{}^2_1$, $\hat{J}{}^2_1$,
$\tilde{J}{}^2_2$ and $\hat{J}{}^2_2$.

The SE equations that govern the six unknowns of
this problem are: 
\begin{eqnarray}
  [1+\delta^{(K)}(1-\epsilon)]
  \left( \begin{array}{l}
      {S^0_0} \\ 
      {S^2_0} \\ 
      {\tilde{S}^2_1} \\
      {\hat{S}^2_1} \\ 
      {\tilde{S}^2_2} \\ 
      {\hat{S}^2_2}
    \end{array} \right) =
    \left( \begin{array}{cccccc}
        0 & 0 & 0 & 0 & 0 & 0 \\
        0 & M_{11} & M_{12} & M_{13} & M_{14} & M_{15} \\ 
        0 & M_{21} & M_{22} & M_{23} & M_{24} & M_{25} \\
        0 & M_{31} & M_{32} & M_{33} & M_{34} & M_{35} \\
        0 & M_{41} & M_{42} & M_{43} & M_{44} & M_{45} \\
        0 & M_{51} & M_{52} & M_{53} & M_{54} & M_{55} 
      \end{array} \right) 
    \left( \begin{array}{l}
        {S^0_0} \\ 
        {S^2_0} \\ 
        {\tilde{S}^2_1} \\
        {\hat{S}^2_1} \\ 
        {\tilde{S}^2_2} \\ 
        {\hat{S}^2_2}
      \end{array} \right) \nonumber 
\end{eqnarray}
\vspace{-0.3in}
\begin{eqnarray}
\hspace{1.7in}
  + \,
  (1-\epsilon) \, w^{(K)}_{J_uJ_l} \, 
  \left( \begin{array}{r}
      J^0_0 \\
      J^2_0 \\
      \tilde{J}{}^{2}_{1} \\
      -\hat{J}{}^{2}_{1} \\
      \tilde{J}{}^{2}_{2} \\
      -\hat{J}{}^{2}_{2}
    \end{array} \right)
  + 
  \epsilon 
  \left( \begin{array}{cccccc}
      B_{\nu_{ul}} \\
      0 \\ 
      0 \\
      0 \\
      0 \\
      0
    \end{array} \right)\,,
\end{eqnarray}
where $S^K_Q=(2h\nu_{ul}^3/c^2) \,[(2J_l+1)/\sqrt{2J_u+1}\,]\,\, \rho^K_Q$.

The $M_{ij}$-quantities of the {\it magnetic operator} ${\cal M}$ are 
coefficients that depend on the strength
and orientation of the local magnetic field
(see Table 1 in Landi Degl'Innocenti {\it et al.}, 1990; for their
explicit values), $w^{(K)}_{J_uJ_l}$ is a numerical factor depending on the
total angular momentum of the levels involved in the transition (see Table I
in Landi Degl'Innocenti, 1984; and note that it is unity for
a $J_l=0$ and $J_u=1$ line transition), 
$\epsilon$ is the collisional destruction probability due to inelastic
collisions, and $B_{\nu_{ul}}$ is the Planck function. We point out that
$\delta^{(K)}$ is the collisional depolarizing rate due to
elastic collisions measured in units of the Einstein $A_{ul}$ coefficient,
with $\delta^{(0)}=0$ in the first equation.

These SE equations have a clear physical meaning. The magnetic operator
${\cal M}$ couples the $K=2$ statistical tensors among them. This is a
local term because its $M_{ij}$ coefficients only depend on the local value
of the magnetic field. The second term in the {\em r.h.s.} of Eq. (12) is the radiative coupling
term. It couples the atomic system with the radiation field and therefore it is highly
non-local. Finally, the third term is the unpolarized thermal source.

Since, as mentioned above, the orientation components are zero
only three Stokes parameters are relevant in this
problem: $I_{x\vec{\Omega}}$, $Q_{x\vec{\Omega}}$ and
$U_{x\vec{\Omega}}$. Due to the fact that 
the lower level is assumed to be unpolarized, 
the radiative transfer equation for each Stokes parameter
is decoupled from the others:
\begin{eqnarray}
  \frac{\rm d}{\rm d \tau_x} {\cal Z}_{x\vec{\Omega}}={\cal
  Z}_{x\vec{\Omega}} - {\cal S}_{\cal Z},
\end{eqnarray}
with ${\cal Z}_{x\vec{\Omega}}$ the Stokes parameter $I_{x\vec{\Omega}}$, 
$Q_{x\vec{\Omega}}$ or $U_{x\vec{\Omega}}$, and d$\tau_x$=$-(\chi_l\phi_x+\chi_c)$d$s$ (with $s$ the geometrical distance along the
ray path and 
$\chi_{l,\, c}$  the line-integrated and continuum opacities). Although
our code is very general, and takes into account the effect
of a background continuum, for notational
simplicity we will not consider it explicitly
in the following equations.
Thus, the line contributions to the source 
functions components $S_{\cal Z}$ are: 
\begin{eqnarray}
  S^{line}_I=S^0_0 + w^{(2)}_{J_uJ_l} \Big{\{}
  \frac{1}{2\sqrt{2}} (3 \mu^2-1)S^2_0 - 
  \sqrt{3} \mu \sqrt{1-\mu^2} (\cos \chi \tilde{S}^2_1 - \sin
  \chi \hat{S}^2_1) \nonumber 
\end{eqnarray}
\vspace{-0.3in}
\begin{eqnarray}
  \hspace{1.9in}+  \frac{\sqrt{3}}{2} (1-\mu^2) (\cos
  2\chi \, \tilde{S}^2_2-\sin 2\chi \, \hat{S}^2_2) \Big{\}},
\end{eqnarray}
\vspace{-0.3in}
\begin{eqnarray}
  S^{line}_Q=w^{(2)}_{J_uJ_l}\Big{\{}\frac{3}{2\sqrt{2}}(\mu^2-1) S^2_0 -
  \sqrt{3}  \mu \sqrt{1-\mu^2} (\cos \chi
  \tilde{S}^2_1 - \sin 
  \chi \hat{S}^2_1) \nonumber 
\end{eqnarray}
\vspace{-0.3in}
\begin{eqnarray}
\hspace{1.9in} - \frac{\sqrt{3}}{2} (1+\mu^2) (\cos
  2\chi \, \tilde{S}^2_2-\sin 2\chi \, \hat{S}^2_2) \Big{\}},
\end{eqnarray}
\vspace{-0.3in}
\begin{eqnarray}
S^{line}_U= w^{(2)}_{J_uJ_l}\sqrt{3} \,\Big{\{} \sqrt{1-\mu^2} ( \sin \chi
  \tilde{S}^2_1+\cos \chi \hat{S}^2_1) \nonumber 
\end{eqnarray}
\vspace{-0.3in}
\begin{eqnarray}
\hspace{1.3in}  +
  \mu (\sin 2\chi \, \tilde{S}^2_2 + \cos 2\chi \, \hat{S}^2_2) \Big{\}},
\end{eqnarray}
where the $S^K_Q$-quantities are given in terms of the density-matrix elements
$\rho^K_Q$ as indicated by the expression given after Eq. (12).

Since the transfer equations (13) are decoupled, 
it is straightforward to write the
formal solution for $I_{x\vec{\Omega}}$, $Q_{x\vec{\Omega}}$ and $U_{x\vec{\Omega}}$, as in the standard unpolarized case, through
the monochromatic $\Lambda_{\nu\Omega}$
operator (see Mihalas, 1978). Thus, with the values of the density matrix
elements we calculate $S_I$, $S_Q$ and $S_U$, and then any suitable
formal solution method of the 
standard RT equation can be used to calculate the Stokes
parameters $I_{x\vec{\Omega}}$, $Q_{x\vec{\Omega}}$ and $U_{x\vec{\Omega}}$ 
at each grid point of the chosen spatial grid of NP points, 
for all the frequencies and directions
of the chosen numerical quadrature.
This allows us to write the radiation field tensors in the absence of
a background continuum as:
\begin{equation}
  \left( \begin{array}{r}
      {\bf J}^0_0 \\
      {\bf J}^2_0 \\
      \tilde{{\bf J}}{}^{2}_{1} \\
      -\hat{{\bf J}}{}^{2}_{1} \\
      \tilde{{\bf J}}{}^{2}_{2} \\
      -\hat{{\bf J}}{}^{2}_{2}
    \end{array} \right)=
    \left( \begin{array}{cccccc}
        {\bf {\Lambda}}_{00} &{\bf {\Lambda}}_{01} &{\bf {\Lambda}}_{02} &{\bf {\Lambda}}_{03} &{\bf {\Lambda}}_{04} &{\bf {\Lambda}}_{05} \\
        {\bf {\Lambda}}_{10} &{\bf {\Lambda}}_{11} &{\bf {\Lambda}}_{12} &{\bf {\Lambda}}_{13} &{\bf {\Lambda}}_{14} &{\bf {\Lambda}}_{15} \\
        {\bf {\Lambda}}_{20} &{\bf {\Lambda}}_{21} &{\bf {\Lambda}}_{22} &{\bf {\Lambda}}_{23} &{\bf {\Lambda}}_{24} &{\bf {\Lambda}}_{25} \\
        {\bf {\Lambda}}_{30} &{\bf {\Lambda}}_{31} &{\bf {\Lambda}}_{32} &{\bf {\Lambda}}_{33} &{\bf {\Lambda}}_{34} &{\bf {\Lambda}}_{35} \\
        {\bf {\Lambda}}_{40} &{\bf {\Lambda}}_{41} &{\bf {\Lambda}}_{42} &{\bf {\Lambda}}_{43} &{\bf {\Lambda}}_{44} &{\bf {\Lambda}}_{45} \\
        {\bf {\Lambda}}_{50} &{\bf {\Lambda}}_{51} &{\bf {\Lambda}}_{52} &{\bf {\Lambda}}_{53} &{\bf {\Lambda}}_{54} &{\bf {\Lambda}}_{55}
      \end{array} \right) 
    \left( \begin{array}{l}
        {{\bf S}^0_0} \\ 
        {{\bf S}^2_0} \\ 
        {\tilde{\bf S}^2_1} \\
        {\hat{\bf S}^2_1} \\ 
        {\tilde{\bf S}^2_2} \\ 
        {\hat{\bf S}^2_2}
      \end{array} \right)+
    \left( \begin{array}{r}
        {{\bf  T}^0_0} \\ 
        {{\bf T}^2_0} \\ 
        {\tilde{{\bf T}}^2_{1}} \\
        {-\hat{{\bf T}}^2_{1}} \\ 
        {\tilde{{\bf T}}^2_{2}} \\ 
        {-\hat{{\bf T}}^2_{2}}
      \end{array} \right),
\end{equation}
where the NP$\times$NP operators ${\bf {\Lambda}}_{\alpha \beta}$ are
frequency and angular 
weighted averages of the
standard $\Lambda_{\nu\Omega}$ operator, and ${\bf J}^K_Q$, ${\bf S}^K_Q$ and
${\bf T}^K_Q$ are vectors of length NP, with the
${\bf T}^K_Q$ components given by expressions similar to 
Eqs. (8)-(11) but using the transmitted Stokes parameters due to the
incident radiation at the boundaries instead of $I_{x\vec{\Omega}}$,
$Q_{x\vec{\Omega}}$ and $U_{x\vec{\Omega}}$.
The analytic expressions of these $\Lambda$-like 
operators can be found in Manso Sainz \& Trujillo Bueno (in preparation).
As shown below, the only operator of relevance for our iterative approach is
${\bf \Lambda}_{00}$, which is nothing but
the $\bar{\Lambda}$ operator of the standard
unpolarized case:
\begin{equation}
  {{\Lambda}}_{00}(i, j)\,=\,\frac{1}{4\pi} \int \phi_x {\rm d}x \oint {\rm d} {\bf
  \Omega} \,\, {{\Lambda}}_{\nu\Omega}(i, j) .
\end{equation}
It can be
demonstrated that in plane-parallel atmospheres all the non-diagonal operators
are zero except ${\bf \Lambda}_{01}={\bf \Lambda}_{10}$ (see also Landi
Degl'Innocenti {\em et al.}, 1990; Nagendra, Frisch \&
Faurobert-Scholl, 1998), and that
in 2D 
media ${\bf \Lambda}_{03}$, ${\bf \Lambda}_{05}$, ${\bf \Lambda}_{13}$, ${\bf \Lambda}_{15}$,
${\bf \Lambda}_{23}$, ${\bf \Lambda}_{25}$, ${\bf \Lambda}_{34}$, ${\bf \Lambda}_{45}$ and their
symmetric ones are zero (Manso Sainz \& Trujillo Bueno, in preparation).

\section{Iterative methods of solution}

The most simple iterative scheme to solve this coupled set of equations is
the $\Lambda$-iteration method. Given an estimate 
${S^K_Q}^{old}$ of the unknowns
(i.e. of the $S^K_Q$ tensors at all the spatial grid-points)
solve formally the transfer equations to
calculate the corresponding six 
${{J}^K_Q}{}^{old}$ radiation field tensors at each spatial grid-point ``$i$''. Then, 
introduce these $J^K_Q{}^{old}(i)$ values into the SE
equations at each spatial grid-point independently
and get an improved $new$ set of $S^K_Q$-values. As shown
by Trujillo Bueno and Manso Sainz (1999), this $\Lambda$-iteration
method can  be used as a reliable solution method {\it if}
one initializes with the self-consistent $\rho^0_0$-values
corresponding to the case in which polarization phenomena are neglected.

In order to develop iterative methods characterized by an extremely
high convergence rate, and that can be applied
to find the self-consistent solution
independently of the chosen initialization, we need to 
account {\it implicitly} for some of the ``new'' values of the
unknowns $\rho^K_Q$-elements. To this end, in the following
we generalize to the Hanle effect regime the
Jacobi-based ALI method of Olson {\it et al.} (1986) and the 
iterative methods of Trujillo Bueno and Fabiani Bendicho (1995)
that are based on Gauss-Seidel iteration (see also Trujillo Bueno and Manso Sainz, 1999).

In order to derive a Jacobi-based ALI scheme we do the same
as with the $\Lambda$-iteration method, except that 
in order to calculate $J^0_0(i)$ we
use the $new$ value of $S^0_0$ (instead of the $old$ one) at the
grid point ``$i$'' being considered. Since this 
``{\it new}'' value is not yet known, we
implicitly write it {\it only} in the expression of $J^0_0$, i.e.
we write
\begin{equation}
  \begin{array}{ll} 
    J^0_0 (i)\,{\approx}\,
    J^0_0{}^{{old}} (i) \,+\,
    {\Lambda_{00}} (i, i) {\delta S^0_0} (i)\,, \\
    {J^2_Q} (i)\,{\approx}\,
    {J^2_Q}{}^{{old}}(i),
  \end{array}
\end{equation}
where $\delta S^0_0(i) = S^0_0{}^{new}(i)-S^0_0{}^{old}(i)$.
After substitution of these
expressions for the $J^K_Q(i)$ quantities 
into the SE equations, we obtain at each grid-point ``$i$''
independently a system of six equations with six unknowns 
that can be solved easily to find
the $new$ values of the six statistical tensors $S^K_Q(i)$.
We point out that this is equivalent to applying the 
operator splitting technique to the
${\bf \Lambda}_{00}$ operator only. It
can be demonstrated that no gain is obtained if the splitting is applied
to the whole set of 36 operators of Eq. (17)
(see next subsection below). 

We emphasize that, at each iterative step, the ${\delta S}^K_Q$ corrections
are made {\em point by point}. Thus, a better idea 
than Jacobi's method would be to apply the method based on
Gauss-Seidel (GS) iteration
of Trujillo Bueno and Fabiani Bendicho (1995), i.e. to  
do the same as with the $\Lambda$-iteration method, except that 
in order to calculate $J^0_0(i)$ we
use the $new$ values of $S^0_0$ (instead of the $old$ ones) at the
grid point ``$i$'' being considered and {\it also} at the grid-points
1,2,...,$i-1$ that have been {\it previously} considered
as we advance from the atmosphere's boundary at which the first
correction was made. Since ${S^0_0}^{new}(i)$ is not yet known, we
implicitly write it {\it only} in the expression of $J^0_0(i)$, i.e.
we write
\begin{equation}
  \begin{array}{ll} 
    J^0_0 (i)\,{\approx}\,
    J^0_0{}^{{old}{\rm and}{new}} (i) \,+\,
    {\Lambda_{00}} (i, i) {\delta S^0_0} (i)\,, \\
    {J^2_Q} (i)\,{\approx}\,
    {J^2_Q}{}^{{old}}(i),
  \end{array}
\end{equation}
where $J^0_0{}^{{old}{\rm and}{new}} (i)$ is the averaged
mean intensity calculated using the ``$new$'' values
of $S^0_0$ at the grid-points 1,2,...,$i-1$ and the ``$old$''
$S^K_Q$-values at points $i,i+1,i+2,...,{\rm NP}$, with
NP the total number of points of the spatial grid (see Trujillo
Bueno and Fabiani Bendicho, 1995; for details).

After substitution of these
expressions for the $J^K_Q(i)$ quantities 
into the SE equations, we obtain at each grid-point ``$i$''
a system of six equations with six unknowns 
that can be easily solved to find
the $new$ values of the statistical tensors $S^K_Q(i)$.
By implementing this GS-based iterative scheme as suggested
in the conclusions of the paper by Trujillo Bueno and Fabiani Bendicho
(1995) the total computational work required to achieve the self-consistent
solution is a factor 4 smaller than with the previous Jacobi-based method.

An extra important improvement of the convergence rate 
can be achieved by multiplying each
$\delta S^K_Q$ GS-correction by a numerical 
factor ($\omega$) lying between 1 and 2.
This is the successive over-relaxation (SOR) method, i.e.

\begin{equation}
{\delta {S^K_Q}}^{\rm SOR}\,=\, \omega {\delta S^K_Q}^{\rm GS}
\end{equation}
The optimal value of $\omega$ can be easily found
(see Trujillo Bueno \& Fabiani Bendicho 1995); however, a good convergence rate can be obtained by just choosing $\omega=1.5$. 

\begin{figure}[t]
  \epsfxsize=8truecm
  \epsfbox[-120 15 384 340]{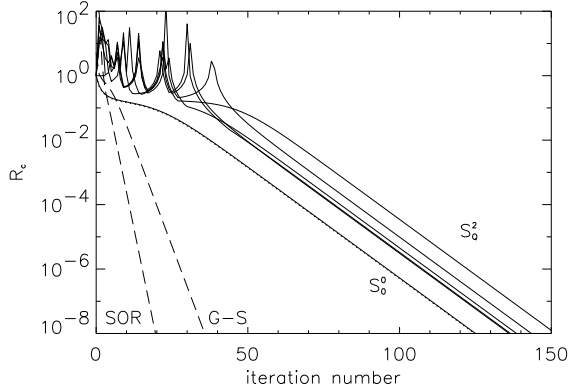}
  \caption{The convergence rate of the six $S^K_Q$ unknowns for
  the Jacobi method (solid lines), and of $S^0_0$ for the Gauss-Seidel
  and the SOR method with $\omega=1.5$
  (dashed lines). See the text for the meaning of the dotted line.}
\end{figure}
Figure 1 shows the convergence rates 
of the six $S^K_Q$ unknowns, i.e. it shows the variation with
the iteration number of the maximum relative change 
($R_c$) in these quantities.
The problem considered was the Hanle effect in a 2D model atmosphere
with $\epsilon=10^{-4}$, $\delta^{(2)}=0$ and
with a non-vertical magnetic field vector. The solid lines refer
to the Jacobi-based ALI method.
We point out that the $S^2_Q$ tensor
elements converge at the same rate as $S^0_0$. The reason for
this is that all the $J^K_Q$
radiation field tensors are dominated by the Stokes I
parameter, and this specific intensity is basically set by the value of
$S^0_0$ (see Trujillo Bueno and Manso Sainz, 1999).
Furthermore, the convergence rate of $S^0_0$ is the same in 
polarized (with or without magnetic fields) and in 
unpolarized problems as shown by the dotted line
in Fig. 1 that cannot be distinguished from the $S^0_0$ solid-line (see
also Faurobert-Scholl {\em et al.}, 1997).
It is important to point out that this is a general behaviour and it is 
independent of the geometry of the medium (1D, 2D or 3D).

The convergence rates for the Gauss-Seidel and SOR iterative methods 
are also plotted in Fig.1. For clarity reasons the convergence rates for
the $S^2_Q$ elements have been omitted because they present
a similar behaviour. As seen in the figure
our GS method is four times faster than
Jacobi, while our SOR method would be a factor 10 if the calculation
had been performed with the optimal $\omega$-value (see also
Trujillo Bueno and Manso Sainz,  1999).

\section{Some illustrative examples}
In the following we show some illustrative examples of the Hanle
effect in 2D atmospheric models
comparing the results with the corresponding 1D case.
Although our code is very general and can deal with
realistic 2D and 3D scenarios with 
horizontal periodic boundary conditions, here we will
restrict ourselves
to sinusoidal fluctuations of the Planck function along the
horizontal X-axis:
\begin{equation}
  B_\nu= \bar{B}_\nu(z) + \Delta B_\nu \cos (k_x x),
\end{equation}
with $k_x=2 \pi / L$ the horizontal wavenumber, and $L$ the 
horizontal wavelength of the thermal inhomogeneities. All
the geometrical distances are measured in units of the opacity scale
height (${\cal H}_\chi \simeq 100$ km). We assume a gaussian
line absorption profile and consider different $k_x$-values,
being $k_x=0$ the plane-parallel 1D limit. 

\begin{figure}[t]
  \epsfxsize=8truecm
  \epsfbox[24 474 359 765]{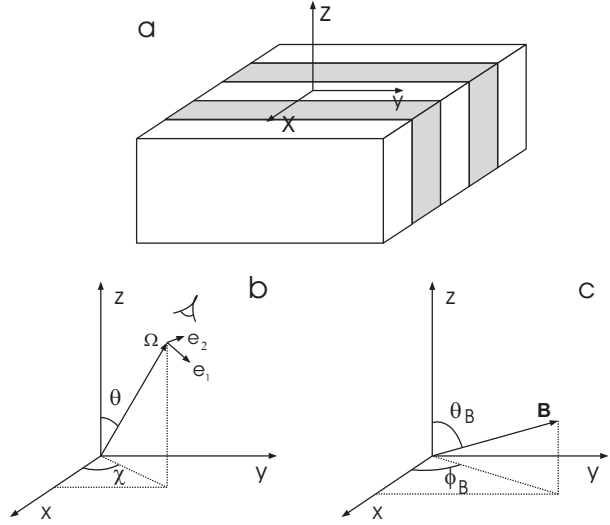}
  \caption{(a) Schematic geometry of a 2D atmosphere with a sinusoidally
  varying 
  temperature inhomogeneity. (b) Show the angles defining the line-of-sight
  ($\vec{\Omega}$) and the polarization unit vectors $\vec{e}_1$
  and $\vec{e}_2$. The positive $Q_{x\vec{\Omega}}$ direction is along
  $\vec{e}_1$. (c) Angles defining the magnetic
  field orientation.}
\end{figure}
Figure 2 is presented to facilitate the understanding of the geometry of
the problem and to indicate the chosen positive and negative directions
of the Stokes Q-parameter.
In Fig. 2$a$ the shaded and white ``slabs'' simply aim at
visualizing the coolest and hottest
regions of the assumed 2D model atmosphere, respectively. 
Note, however, that the chosen 2D medium
is not composed of such embedded slabs, since it is
characterized by the horizontal
temperature fluctuations given by the previous equation.
These ``slabs'' should be understood as 
infinite along the Y-direction.
Fig. 2$b$ shows the angles that specify the direction of propagation
of the ray under consideration, while Fig. 2c gives the 
angles that determine
the orientation of the magnetic field vector.
As seen in Fig. 2$a$ the Y-Z plane at X=0 is dividing 
the hottest ``slabs'' in two equal halves. 
We will show the emergent polarization profiles 
for simulated observations made along the 
hottest ``slabs'' (i.e. at 
the horizontal position X=0 and for a line of sight
with $\chi=90^0$). Thus, the
positive direction of the Stokes Q-parameter lies along the slabs, and the
negative one is perpendicular to them.

\begin{figure}[t]
  \epsfxsize=8truecm
  \epsfbox[-120 15 384 340]{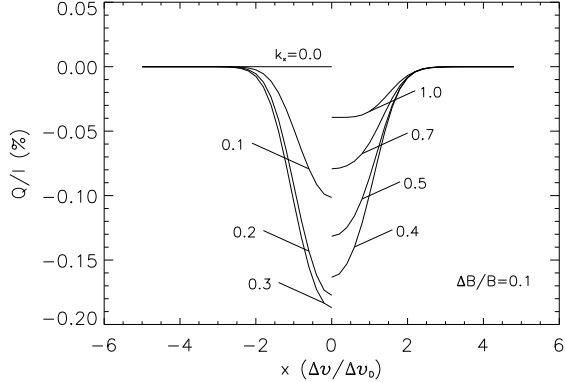}
  \caption{Emergent Q/I profiles at disk center ($\mu=1$) for different
  horizontal wavenumber $k_x$-values. Here $\Delta B_\nu / \bar{B}_\nu =0.1$.}
\end{figure}
In Fig. 3 we consider the case of resonance line polarization in
this 2D atmosphere. 
When observing at disk center (i.e. at $\mu={\rm cos}\theta=1$) we find that 
there is no polarization signal corresponding to the
plane-parallel 1D limit (i.e. for $k_x\,=\,0$),
as it must indeed be the case because the radiation field 
of an unmagnetized 1D medium has axial symmetry,
and the radiation field tensors ${\bar J}^2_1$ and ${\bar J}^2_2$
are then zero. As we increase the horizontal wavenumber $k_x$ 
(i.e. the parameter that 
measures the degree of horizontal inhomogeneity of our 2D model)
the radiation field looses its axial symmetry, 
and the polarization signal $Q/I$ starts to increase accordingly.
However, when the horizontal temperature inhomogeneities 
are smaller than about 2000 km (i.e. for wavenumbers $k_x>0.3$), the
polarization signal decreases as
the wavenumber $k_x$ further increases. This is because we start 
then to approach
the limiting case of an atmosphere composed of optically thin
irregularities, for which the radiation field recovers
the axial symmetry characteristic of a 1D medium.

\begin{figure}[t]
  \epsfxsize=13truecm
  \epsfbox[20 10 520 562]{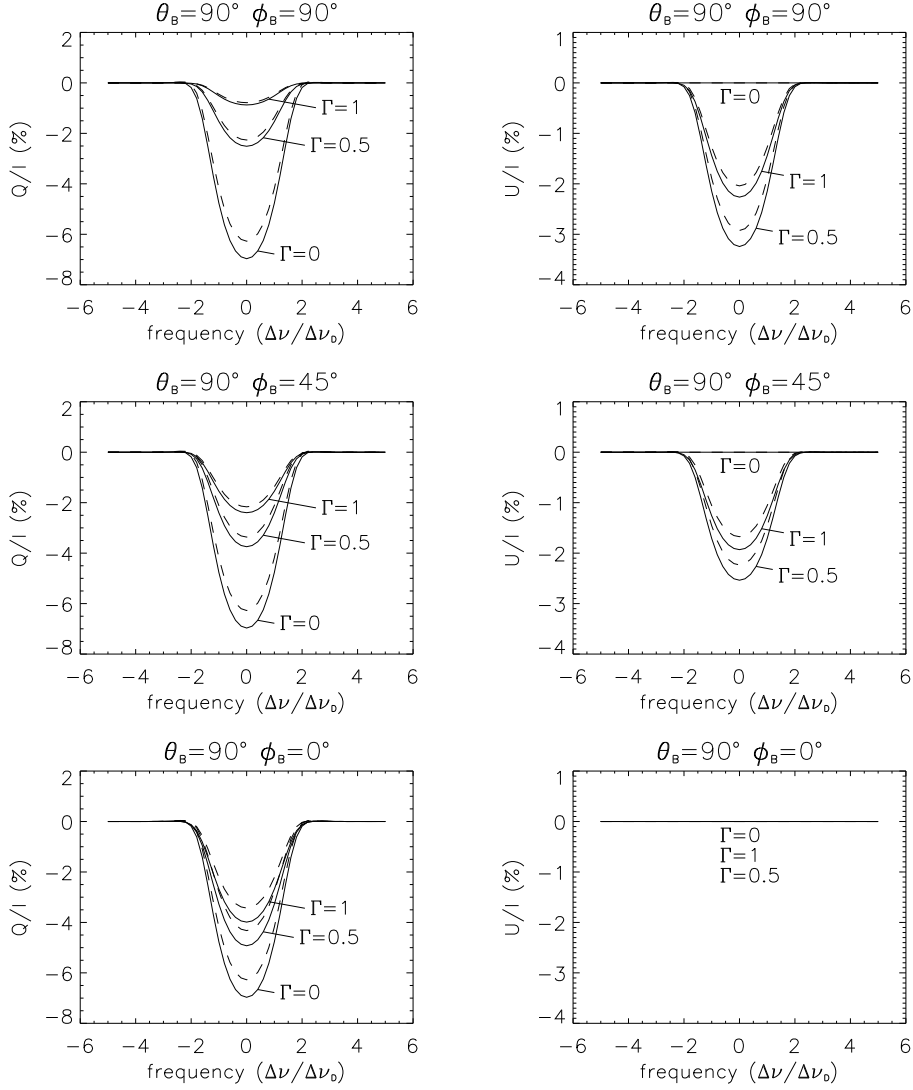}
  \caption{Emergent Q/I and U/I profiles in 
  the assumed 2D atmosphere (solid lines)
  with wavenumber $k_x=0.3$ and $\Delta B_\nu/\bar{B}_\nu=0.1$.
  The line of sight observation is at $\mu=0.1$ and $\chi=90^o$.
  Dashed lines indicate the 1D case.}
\end{figure}
Fig. 4 shows the emergent $Q/I$ and $U/I$ profiles  
for simulated high-spatial resolution observations
with a line of sight having $\chi=90^o$ and made close to
the solar limb ($\mu=0.1$) in a 2D atmosphere with 
$\Delta B_\nu/\bar{B}_\nu=0.1$ and $k_x=0.3$ (i.e. with $L\,{\approx}\,2000$ km). 
The magnetic field direction is defined through the azimuthal and polar
angles $\phi_B$ and $\theta_B$ (see Fig. 2$c$), and its strength through the
parameter $\Gamma=0.88 g_J B/A_{ul}$, with $g_J$ the upper level Land\'e factor, $B$ the
magnetic field in Gauss, and $A_{ul}$ the spontaneous emission Einstein
coefficient measured in units of $10^{7}s^{-1}$. Since the line-of-sight  
lies along the slabs (i.e. $\chi=90^o$), 
we find that in the absence of magnetic field (i.e. for $\Gamma=0$)
$U/I$ is zero due to
symmetry reasons. The two uppermost panels show how a magnetic
field parallel to the slab (i.e. with $\phi_B=\theta_B=90^o$) 
substantially changes
the emergent polarization signal. However, when the magnetic field is
perpendicular to the slab (see the lowermost panels
with $\theta_B=90^o$ and $\phi_B=0^0$), the change with $\Gamma$ of $Q/I$ is
the smallest one and
$U/I$ always remains zero. The two central panels show the
intermediate case with $\phi_B=45^o$. 
The dashed lines show the results for the plane-parallel $k_x=0$ case.

\begin{figure}[t]
  \epsfxsize=8truecm
  \epsfbox[-120 15 384 340]{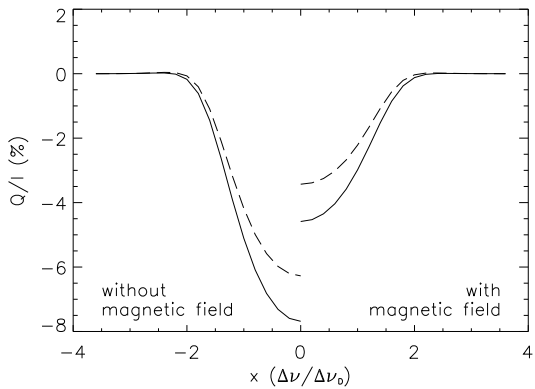}
  \caption{As Fig. 4 but taking $\Delta B_\nu/\bar{B}_\nu=0.2$ and
  including the horizontal
  opacity inhomogeneities given by Eq. (23).
  The magnetic field has $\Gamma=1,
  \theta_B=90^o$, and  $\phi=0^o$.
   Solid lines show the 2D case, and dashed-lines the 1D one.}
\end{figure}
Fig. 5 shows the emergent fractional linear polarization $Q/I$ at $\mu=0.1$
and $\chi=90^o$, when the Planck function varies sinusoidally 
(with $k_x=0.3$ and $\Delta B_\nu/\bar{B}_\nu=0.2$), and there exists 
additionally a horizontal
fluctuation in the line opacity $\chi_l$
that is anticorrelated with that of the Planck function, i.e.
\begin{equation}
  \chi_l = \bar{\chi}_l [1+\alpha \cos (k_x x)],
\end{equation}
with $k_x=0.3$ and $\alpha=-0.2$.
The curves situated on the left-hand-side of the figure
show the 1D and 2D results corresponding
to the zero magnetic field case, while the curves on the {\it r.h.s.}
correspond to a case of a magnetic field with $\Gamma=1$, $\theta_B=90^o$,
and $\phi_B=0^o$. 
For this magnetic and line-of-sight geometry $U/I=0$. It can be seen that,
for this particular model atmosphere and geometry, a Hanle-effect
diagnostic of very high spatial resolution observations would lead
to an underestimation the magnetic field strength for interpretations
based on the plane-parallel 1D approximation.

\section{Conclusions}

We have developed a Hanle effect code that
allows the numerical simulation of resonance line polarization
signals in the presence of weak magnetic fields
in 1D, 2D and 3D media.
The governing equations have been formulated 
working within the framework of the density matrix 
polarization transfer theory of Landi Degl'Innocenti (1983, 1985).
These SE and RT equations are the same independently
of whether we are considering 1D, 2D or 3D atmospheric models.
The six $\rho^K_Q$-unknowns of the problem are
neither frequency nor angle dependent, since they only vary with the spatial
position.

Three different iterative schemes 
that were originally developed for RT applications
in the unpolarized case
have been generalized to solve this 
set of equations: Jacobi (ALI), Gauss-Seidel and SOR 
(Trujillo Bueno \& Fabiani Bendicho, 1995, 
Trujillo Bueno \& Manso Sainz 1999). This kind of 
iterative methods does not
make use of any matrix inversion, and essentially maintain the
$\Lambda$-iteration simplicity. The only difference between
the 1D, 2D and 3D versions of our Hanle effect code lies in the
formal solution routine that calculates the radiation field tensors
from the current values of the density-matrix elements. To this end,
in 2D we use the formal solver developed by Auer, Fabiani Bendicho
and Trujillo Bueno (1994) and in 3D 
we use the one presented at this workshop
by Fabiani Bendicho and Trujillo Bueno (1999). 

We have also shown some Hanle effect 
results for 1D and 2D media. These calculations 
illustrate how weak magnetic fields and horizontal radiative transfer effects
compete to modify the scattering line polarization signals expected from
plane-parallel 1D atmospheres. Thus, further careful investigations must be
done in order to separate both effects, when diagnosing weak 
solar magnetic fields via the Hanle effect. This type of future studies
should be done thinking in the interpretation of {\it low} spatial
resolution scattering line polarization observations. 
Our numerical approach is very efficient
and suitable to investigate scattering polarization signals for
a variety of atmospheric models having
any desired temperature, density and magnetic field vector 
variations. Another useful research
that can be done with our Hanle effect codes 
concerns the simulation of polarization signals
emerging from realistic MHD and semi-empirical 2D and 3D models.

\acknowledgements
We thank Egidio Landi Degl'Innocenti for his careful reading
of our paper. Partial support by the Spanish DGES through
project PB 95 -0028 is gratefully acknowledged.


\begin{thebibliography}{}
\bibitem[]{}
Auer, L.\,H., Fabiani Bendicho, P. and Trujillo Bueno, J. (1994),
{\it Astron. Astrophys.},
{\bf 292}, 599
\bibitem[]{}
Bommier, V., Sahal-Br\'echot (1978), {\it Astron. Astrophys.}, {\bf 69}, 57
\bibitem[]{}
Faurobert-Scholl, M., Frisch, H., Nagendra, K.N. (1997), {\it
  Astron. Astrophys.}, {\bf 322}, 896
\bibitem[]{}
Hanle, W. (1924), {\it Z. Phys.}, {\bf 30}, 93
\bibitem[]{}
Landi Degl'Innocenti, E. (1983), {\it Solar Phys.}, {\bf 85}, 3
\bibitem[]{}
Landi Degl'Innocenti, E. (1984), {\it Solar Phys.}, {\bf 91}, 1
\bibitem[]{}
Landi Degl'Innocenti, E. (1985), {\it Solar Phys.}, {\bf 102}, 1
\bibitem[]{}
Landi Degl'Innocenti, E., Bommier, V., Sahal-Br\'echot, S. (1990), {\it Astron. Astrophys.}, {\bf 235}, 459
\bibitem[]{}
Messiah, A., (1969), {\it Quantum Mechanics}, Dunod, Paris
\bibitem[]{}
Mihalas, D. (1978) {\it Stellar Atmospheres}. W.H. Freeman, San Francisco.
\bibitem[]{}
Nagendra, K.N., Frisch, H., Faurobert-Scholl, M. (1998), {\it
  Astron. Astrophys.}, {\bf 332}, 610
\bibitem[]{}
Olson, G. L., Auer, L. H., \& Buchler, J. R. (1986), 
{\it J. Quant. Spectrosc. Radiat. Transfer}, {\bf 35}, 431
\bibitem[]{}
S\'anchez Almeida, J. (1999), {\it These Proceedings}
\bibitem[]{}
Stenflo, J.O. (1994) {\it Solar Magnetic Fields. Polarized Radiation
  Diagnostics}. Kluwer Academic Publishers, Dordrecht.
\bibitem[]{}
Trujillo Bueno, J. (1999), in {\it Solar Polarization}, edited by K.N. Nagendra \& J.O. Stenflo. Kluwer Academic Publishers, 1999. (Astrophysics and Space Science Library ; V. 243), p. 73-96

\bibitem[]{}
Trujillo Bueno, J., Fabiani Bendicho, P. (1995), {\it Astrophys. J.}, {\bf
  455}, 646
\bibitem[]{}
Trujillo Bueno, J., Manso Sainz, R. (1999), {\it Astrophys. J.}, {\bf 516}, 436

\end{thebibliography}
\end{document}